\documentclass[10pt, twocolumn, conference]{IEEEtran}
\usepackage{epsfig, psfrag, amsmath, amsfonts, amssymb, eucal, balance}
\newtheorem{thm}{Theorem}

\newtheorem{lem}{Lemma}

\newtheorem{defn}{Definition}

\DeclareMathAlphabet{\eurm}{U}{eur}{m}{n}
\DeclareMathAlphabet{\mathbsf}{OT1}{cmss}{bx}{n}
\DeclareMathAlphabet{\mathssf}{OT1}{cmss}{m}{sl}
\DeclareMathAlphabet{\mathcsf}{OT1}{cmss}{sbc}{n}

\newcommand{\randomvalue}[1]{\eurm{\uppercase{#1}}}

\DeclareSymbolFont{bsfletters}{OT1}{cmss}{bx}{n}  
\DeclareSymbolFont{ssfletters}{OT1}{cmss}{m}{n}
\DeclareMathSymbol{\bsfGamma}{0}{bsfletters}{'000}
\DeclareMathSymbol{\ssfGamma}{0}{ssfletters}{'000}
\DeclareMathSymbol{\bsfDelta}{0}{bsfletters}{'001}
\DeclareMathSymbol{\ssfDelta}{0}{ssfletters}{'001}
\DeclareMathSymbol{\bsfTheta}{0}{bsfletters}{'002}
\DeclareMathSymbol{\ssfTheta}{0}{ssfletters}{'002}
\DeclareMathSymbol{\bsfLambda}{0}{bsfletters}{'003}
\DeclareMathSymbol{\ssfLambda}{0}{ssfletters}{'003}
\DeclareMathSymbol{\bsfXi}{0}{bsfletters}{'004}
\DeclareMathSymbol{\ssfXi}{0}{ssfletters}{'004}
\DeclareMathSymbol{\bsfPi}{0}{bsfletters}{'005}
\DeclareMathSymbol{\ssfPi}{0}{ssfletters}{'005}
\DeclareMathSymbol{\bsfSigma}{0}{bsfletters}{'006}
\DeclareMathSymbol{\ssfSigma}{0}{ssfletters}{'006}
\DeclareMathSymbol{\bsfUpsilon}{0}{bsfletters}{'007}
\DeclareMathSymbol{\ssfUpsilon}{0}{ssfletters}{'007}
\DeclareMathSymbol{\bsfPhi}{0}{bsfletters}{'010}
\DeclareMathSymbol{\ssfPhi}{0}{ssfletters}{'010}
\DeclareMathSymbol{\bsfPsi}{0}{bsfletters}{'011}
\DeclareMathSymbol{\ssfPsi}{0}{ssfletters}{'011}
\DeclareMathSymbol{\bsfOmega}{0}{bsfletters}{'012}
\DeclareMathSymbol{\ssfOmega}{0}{ssfletters}{'012}

\newcommand{\rvM}{{\randomvalue{M}}}	

\newcommand{\rvm}{{\randomvalue{m}}}	

\newcommand{\rvs}{{\randomvalue{s}}}	

\newcommand{\rvx}{{\randomvalue{x}}}	

\newcommand{\rvy}{{\randomvalue{y}}}	

\begin{document}

\title{A Constrained Channel Coding Approach to Joint Communication and Channel Estimation}
\author{\authorblockN{Wenyi Zhang, Satish Vedantam, and Urbashi Mitra}
\authorblockA{Ming Hsieh Department of Electrical Engineering\\
University of Southern California\\
\{wenyizha, vedantam, ubli\}@usc.edu}
}

\maketitle

\begin{abstract}
A joint communication and channel state estimation problem is investigated, in which reliable information transmission over a noisy channel, and high-fidelity estimation of the channel state, are simultaneously sought. The tradeoff between the achievable information rate and the estimation distortion is quantified by formulating the problem as a constrained channel coding problem, and the resulting capacity-distortion function characterizes the fundamental limit of the joint communication and channel estimation problem. The analytical results are illustrated through case studies, and further issues such as multiple cost constraints, channel uncertainty, and capacity per unit distortion are also briefly discussed.
\end{abstract}

\section{Introduction}
\label{sec:intro}

In this paper, we consider the problem of joint communication and channel estimation over a channel with a time-varying channel state. We consider a noisy channel with a random channel state that evolves with time, in a memoryless fashion, and is neither available to the transmitter nor the receiver. The objective is to have the receiver recover both the information transmitted from the transmitter as well the state of the channel over which the information was transmitted. The problem setting may prove relevant for situations such as environment monitoring in sensor networks \cite{szewczyk04:cacm}, underwater acoustic applications \cite{stojanovic96:joe}, and cognitive radio \cite{haykin05:jsac}. A distinct feature of our problem formulation is that both communication and channel estimation are required.

The interplay between information measures and estimation (minimum mean-squared error (MMSE) in particular) has long been investigated; see, {\it e.g.}, \cite{guo05:it} and references therein. Previously, however, estimation was only to facilitate information transmission, rather than a separate goal. For example, a common strategy in block interference channels \cite{mceliece84:it} is channel estimation via training \cite{hassibi03:it}. The purpose of channel training is only to increase the information rate for communication, and thus the quality of channel estimate is not traded off with the information rate, as we consider in this paper.

The problem formulation in \cite{sutivong05:it, cover07:arxiv} bears some similarity to the one we consider in that the receiver is interested in both communication and channel estimation. It differs from our work in a critical way: the channel state is assumed non-causally known at the transmitter. In contrast, neither the transmitter nor the receiver knows the channel state in our problem formulation.

Intuitively, there exists a tradeoff between a channel's capability to transfer information and its capability to exhibit state. Increasing randomness in channel inputs increases information transfer while reducing the receiver's ability to estimate the channel. In contrast, deterministic signaling facilitates channel estimation at the expense of zero information transfer. In this paper, we show that the optimal tradeoff can be formulated as a channel coding problem, with the channel input distribution constrained by an average ``estimation cost'' constraint.

The rest of this paper is organized as follows. Section \ref{sec:model} introduces the channel model and the capacity-distortion function, and Section \ref{sec:formulation} formulates the equivalent constrained channel coding problem. Section \ref{sec:example} illustrates the application of the capacity-distortion function through several simple examples. Section \ref{sec:issue} briefly discusses some related issues including multiple cost constraints, channel uncertainty, and capacity per unit distortion. Finally, Section \ref{sec:conclusion} concludes the paper.

\section{Channel Model}
\label{sec:model}

\begin{figure}[ht]
\psfrag{message}{{\tiny $\rvM$}}
\psfrag{messageest}{{\tiny $\hat{\rvM}$}}
\psfrag{input}{{\tiny $\rvx_i$}}
\psfrag{output}{{\tiny $\rvy_i$}}
\psfrag{state}{{\tiny $\rvs_i$}}
\psfrag{stateest}{\tiny $\hat{\rvs}_i$}
\psfrag{transP}{\tiny $P(y|x, s)$}
\epsfxsize=3.5in
\epsfclipon
\centerline{\epsffile{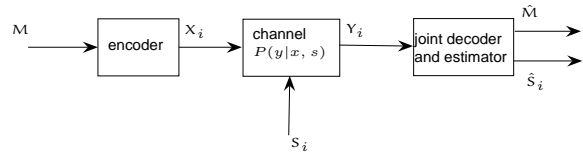}}
\caption{Channel model for joint communication and channel estimation.}
\label{fig:model}
\end{figure}

We consider the channel model in Figure \ref{fig:model}. For a length-$n$ block of channel inputs, a message $\rvm$ is equally probably selected among $\{1, \ldots, \left\lceil e^{nR} \right\rceil\}$, and is encoded by the encoder, generating the corresponding channel inputs $\{\rvx_1, \ldots, \rvx_n\}$. We provide the following definition.

\begin{defn}(Encoder)
An encoder is defined by a function, $f_n: \mathcal{M} = \{1, \ldots, \left\lceil e^{nR} \right\rceil\} \rightarrow \mathcal{X}^n$, for each $n \in \mathbb{N}$.
\end{defn}

The channel is described by a transition function $P(y|x, s)$, which is the probability distribution of the channel output $\rvy$, conditioned on the channel input $\rvx$ and the channel state $\rvs$. Upon receiving the length-$n$ block of channel outputs, the joint decoder and estimator (defined below) declares $\hat{M} \in \{1, \ldots, \left\lceil e^{nR} \right\rceil\}$ as the decoded message, and a length-$n$ block of estimates of the channel state.

For technical purposes, in this paper, we assume that the random channel state evolves with time in a memoryless fashion. We note that this model encompasses the block interference channel model, because we can treat a block as a super-symbol and thus convert a block interference channel into a memoryless channel.

\begin{defn}(Joint decoder and estimator)
A joint decoder and estimator is defined by a pair of functions, $g_n: \mathcal{Y}^n \rightarrow \mathcal{M}$ and $h_n: \mathcal{Y}^n \rightarrow \mathcal{S}^n$, for each $n \in \mathbb{N}$.
\end{defn}

This definition differs from that of the conventional channel decoder ({\it e.g.}, \cite{cover91:book}) in that it explicitly requires estimation of the channel state $\rvs$ at the receiver. The quality of estimation is measured by the distortion function $d: \mathcal{S} \times \mathcal{S} \rightarrow \mathbb{R}^+\cup \{0\}$. That is, if $\hat{\rvs}_i$ is the $i$th element of $h_n(\rvy^n)$, then $d(\rvs_i, \hat{\rvs}_i)$ denotes the distortion at time $i$, $i = 1, \ldots, n$. For technical convenience, we assume that $d(\cdot, \cdot)$ is bounded from above so that there exists a finite $T > 0$ with $d(s, s^\prime) \leq T < \infty$ for any $s, s^\prime \in \mathcal{S}$. Note that for length-$n$ block coding schemes, the average distortion is given by
\begin{eqnarray}
\bar{d}(\rvs^n, \hat{\rvs}^n) = \frac{1}{n}\sum_{i = 1}^n d(\rvs_i, \hat{\rvs}_i).
\end{eqnarray}

Finally, we have the following definitions.

\begin{defn}(Achievable rate)
\label{defn:ach-rate}
A nonnegative number $R(D)$ is an achievable rate if there exist a sequence of encoders and corresponding joint decoders and estimators such that (a) the average probability of decoding error $P_e^{(n)} = (1/\left\lceil e^{nR(D)} \right\rceil)\cdot \sum_{m = 1}^{\left\lceil e^{nR(D)} \right\rceil} \mathrm{Pr}[\hat{\rvm} \neq m|\rvm = m]$ tends to zero as $n \rightarrow \infty$; and (b) the average distortion in channel state estimation,
\begin{eqnarray}
\label{eqn:ave-dist}
\limsup_{n \rightarrow \infty} \mathbf{E} \bar{d}(\rvs^n, \hat{\rvs}^n) \leq D.
\end{eqnarray}
\end{defn}

\begin{defn}(Capacity-distortion function)
\label{defn:cap-dist}
The capacity-distortion function is defined as
\begin{eqnarray}
C(D) = \sup_{f_n, g_n, h_n} R(D).
\end{eqnarray}
\end{defn}

{\it Remark}: The reader may want to distinguish between the capacity-distortion function and the rate-distortion function in lossy source coding \cite{cover91:book}. The capacity-distortion function is defined with respect to a state-dependent channel, seeking to characterize the fundamental tradeoff between the rate of information transmission and the distortion of state estimation. In contrast, the rate-distortion function is defined with respect to a source distribution, seeking to characterize the fundamental tradeoff between the rate of its lossy description and the achievable distortion due to the description.

\section{A Constrained Channel Coding Formulation}
\label{sec:formulation}

In this section, we show that the joint communication and channel estimation problem can be equivalently formulated as a constrained channel coding problem. For this purpose, the following minimum conditional distortion will be important. The minimum conditional distortion function is defined for each possible realization of the channel input $\rvx$, as
\begin{eqnarray}
\label{eqn:d-ast}
d^\ast(x) = \inf_{h_0: \mathcal{X} \times \mathcal{Y} \rightarrow \mathcal{S}} \mathbf{E}\left[d(\rvs, h_0(x, \rvy))\right],
\end{eqnarray}
where the expectation is with respect to the channel state $\rvs$ and the channel output $\rvy$ conditioned upon the channel input $\rvx = x$, and $h_0: \mathcal{X} \times \mathcal{Y} \rightarrow \mathcal{S}$ denotes an arbitrary one-shot estimator of $\rvs$ given the channel input and output.

The following theorem establishes the constrained channel coding formulation.
\begin{thm}
\label{thm:concap}
The capacity-distortion function for the channel model in Figure \ref{fig:model} is given by
\begin{eqnarray}
C(D) = \sup_{P_\rvx \in \mathcal{P}_D} I(\rvx; \rvy),
\end{eqnarray}
where
\begin{eqnarray}
\label{eqn:pd-defn}
\mathcal{P}_D = \left\{P_\rvx: \sum_{x \in \mathcal{X}} P_\rvx(x) d^\ast(x) \leq D \right\}.
\end{eqnarray}
\end{thm}

{\it Remark}: Theorem \ref{thm:concap} applies to general input/output/state alphabets. If $\rvx$ is a continuous random variable, the summation in (\ref{eqn:pd-defn}) should be understood as an integral over $\mathcal{X}$.

In order to prove Theorem \ref{thm:concap}, we shall employ the following lemmas.

\begin{lem}
\label{lem:knowledge}
For any $(f_n, g_n, h_n)$-sequence that achieves $C(D)$, as $n \rightarrow \infty$, the achieved average distortion (\ref{eqn:ave-dist}) is (in probability) equal to the average distortion with $\hat{\rvs}^n$ replaced by
\begin{eqnarray}
\label{eqn:h-both}
\hat{\rvs}^n = h^\ast_n(\rvx^n, \rvy^n),
\end{eqnarray}
where $h^\ast_n(\rvx^n, \rvy^n)$ denotes the block-$n$ estimator that achieves the minimum average distortion conditioned upon both the block-$n$ channel inputs and outputs.
\end{lem}
{\it Proof}: For each $n$, let us replace the estimator $h_n$ by $h^\ast_n$ in (\ref{eqn:h-both}), with its first argument being the channel inputs $\hat{\rvx}^n$ corresponding to the decoded message $\hat{\rvM}$. When $\hat{\rvM} = \rvM$, the minimum average distortion is achieved by $h^\ast_n$; when $\hat{\rvM} \neq \rvM$, the increment in the average distortion due to replacing $h_n$ by $h^\ast_n$ is bounded from above because $d(\cdot, \cdot) \leq T < \infty$. By Definitions \ref{defn:ach-rate} and \ref{defn:cap-dist}, as $n \rightarrow \infty$, the average probability of decoding error $P_e^{(n)} \rightarrow 0$. Hence as $n \rightarrow \infty$, the minimum average distortion is achieved by $h^\ast_n(\hat{\rvx}^n, \rvy^n)$, which is further equal to (\ref{eqn:h-both}), in probability. {\bf Q.E.D.}

Lemma \ref{lem:knowledge} shows that the joint decoder and estimator can utilize the reliably decoded channel inputs for channel state estimation. The next lemma, Lemma \ref{lem:decomp}, further shows that the length-$n$ block estimator can be decomposed into $n$ one-shot estimators, each for one channel use.

\begin{lem}
\label{lem:decomp}
For any $(f_n, g_n, h_n)$-sequence that achieves $C(D)$, as $n \rightarrow \infty$, the achieved average distortion (\ref{eqn:ave-dist}) is (in probability) equal to that achieved by
\begin{eqnarray}
\label{eqn:decomp}
\hat{\rvs}_i = h^\ast_0(\rvx_i, \rvy_i),\quad i = 1, \ldots, n,
\end{eqnarray}
where $h^\ast_0(\rvx_i, \rvy_i)$ denotes the one-shot estimator that achieves the minimum expected distortion for $\rvs_i$ conditioned upon both the channel input $\rvx_i$ and output $\rvy_i$.
\end{lem}
{\it Proof}: From Lemma \ref{lem:knowledge}, as $n \rightarrow \infty$, $h_n(\rvy^n)$ is in probability equivalent to $h^\ast_n(\rvx^n, \rvy^n)$. The decomposition (\ref{eqn:decomp}) then follows because the channel is memoryless. For each fixed $n$, we have
\begin{eqnarray}
&&P(\rvs^n|\rvx^n, \rvy^n) = \frac{P(\rvx^n, \rvy^n, \rvs^n)}{P(\rvx^n, \rvy^n)}\nonumber\\
&=& \frac{\prod_{i = 1}^n P(\rvs_i, \rvx_i, \rvy_i)}{\sum_{\rvs^n} \prod_{i = 1}^n P(\rvs_i, \rvx_i, \rvy_i)}\nonumber\\
&=& \frac{\prod_{i = 1}^n P(\rvs_i, \rvx_i, \rvy_i)}{\prod_{i = 1}^n \left[\sum_{\rvs_i} P(\rvy_i|\rvx_i, \rvs_i) P(\rvs_i)\right] P_\rvx(\rvx_i)}\nonumber\\
&=& \prod_{i = 1}^n \frac{P(\rvs_i, \rvx_i, \rvy_i)}{P(\rvy_i|\rvx_i) P_\rvx(\rvx_i)} = \prod_{i = 1}^n P(\rvs_i |\rvx_i, \rvy_i).
\end{eqnarray}
As we take $n \rightarrow \infty$, the lemma is established. {\bf Q.E.D.}

{\it Proof of Theorem \ref{thm:concap}}: From Lemmas \ref{lem:knowledge} and \ref{lem:decomp}, we can rewrite the average distortion constraint (\ref{eqn:ave-dist}) as
\begin{eqnarray}
\label{eqn:proof-1}
&&\limsup_{n \rightarrow \infty} \frac{1}{n} \sum_{i = 1}^n \mathbf{E} d(\rvs_i, \hat{\rvs}_i) \leq D\nonumber\\
&\Rightarrow& \limsup_{n \rightarrow \infty} \frac{1}{n} \sum_{i = 1}^n \mathbf{E} d(\rvs_i, h^\ast_0(\rvx_i, \rvy_i)) \leq D.
\end{eqnarray}
Utilizing (\ref{eqn:d-ast}) and the fact that the channel is memoryless, we can further deduce from (\ref{eqn:proof-1}) that
\begin{eqnarray}
\label{eqn:cost-constraint}
\mathbf{E} d^\ast(X) \leq D.
\end{eqnarray}

So now the constraints in Definition \ref{defn:ach-rate} reduce to having $P_e^{(n)} \rightarrow 0$ as $n \rightarrow \infty$, subject to the constraint (\ref{eqn:cost-constraint}). This is exactly the problem of channel coding with a cost constraint on the input distribution, and Theorem \ref{thm:concap} directly follows from standard proofs; see, {\it e.g.}, \cite{gallager68:book}. {\bf Q.E.D.}

{\it Discussion}:

(1) The proof of Theorem \ref{thm:concap} suggests the joint decoder and estimator first decode the transmitted message in a ``non-coherent'' fashion, then utilize the reconstructed channel inputs along with the channel outputs to estimate the channel states. As the coding block length grows large, such a two-stage procedure becomes asymptotically optimal.

(2) For each $x \in \mathcal{X}$, $d^\ast(x)$ quantifies its associated minimum distortion. Alternatively, $d^\ast(x)$ can be viewed as the ``estimation cost'' due to signaling with $x$. Hence the average distortion constraint in (\ref{eqn:pd-defn}) regulates the input distribution such that the signaling is estimation-efficient. We emphasize that, $d^\ast(x)$ is dependent on the channel through the distribution of the channel state $\rvs$, and thus differs from other usual costs such as symbol energies or time durations.

(3) A key condition that leads to the constrained channel coding formulation is that the channel is memoryless. Due to the memoryless property, we can decompose a block estimator into multiple one-shot estimators, without loss of optimality asymptotically. If the channel state evolves with time in a correlated fashion, then such a decomposition is generally suboptimal.

\section{Illustrative Examples}
\label{sec:example}

In this section, we discuss several simple examples to illustrate the application of Theorem \ref{thm:concap}.

\subsection{Uniform Estimation Costs}

A special case is that $d^\ast(x) = d_0$ for all $x \in \mathcal{X}$. For such type of channels, the average cost constraint in (\ref{eqn:pd-defn}) exhibits a singular behavior. If $D < d_0$, then the joint communication and channel estimation problem is infeasible; otherwise, $\mathcal{P}_D$ consists of all possible input distributions, and thus the capacity-distortion function $C(D)$ is equal to the unconstrained capacity of the channel. One of the simplest channels with uniform estimation costs is the additive channel $\rvy_i = \rvx_i + \rvs_i$, for which as the receiver reliably decodes $\rvM$, it can subtract off $\rvx_i$ from $\rvy_i$.

\subsection{A Scalar Multiplicative Channel}
\label{subsec:smultiply}

Consider the following scalar multiplicative channel
\begin{eqnarray}
\rvy_i = \rvs_i \rvx_i,
\end{eqnarray}
where all the alphabets are binary, $\mathcal{X} = \mathcal{Y} = \mathcal{S} = \{0, 1\}$, and the multiplication is in the conventional sense for real numbers. The reader may interpret $\rvs$ as the status of an informed jamming source, a fading level, or the status of another transmitter. Activating $\rvs$ to its ``effective status'' $\rvs = 0$ shuts down the link between $\rvx$ and $\rvy$; otherwise, the link $\rvx \rightarrow \rvy$ is essentially noiseless. We take the distortion measure as the Hamming distance: $d(s, \hat{s}) = 1$ if and only if $\hat{s} \neq s$ and zero otherwise.

The tradeoff between communication and channel estimation is straightforward to observe from the nature of the channel: for good estimation of $\rvs$, we want $\rvx = 1$ as often as possible, whereas this would reduce the achieved information rate. In this example, we assume that $P(\rvs = 1) = r \leq 1/2$. We shall optimize $P(\rvx = 1)$, denoted by $p \in [0, 1]$. The channel mutual information is $I(\rvx; \rvy) = H_2(pr) - p\cdot H_2(r)$, where $H_2(\cdot)$ denotes the binary entropy function $H_2(t) = -t \log t - (1 - t)\log(1 - t)$. For $x = 0$, the optimal one-shot estimator is $\hat{\rvs} = 0$ (note that $P(\rvs = 1) = r \leq 1/2$), and the resulting minimum conditional distortion is $d^\ast(0) = r$. For $x = 1$, the optimal one-shot estimator is $\hat{\rvs} = \rvy = \rvs$, leading to $d^\ast(1) = 0$. Therefore the input distribution should satisfy $(1 - p) r \leq D$.

\begin{figure}[ht]
\psfrag{xlabel}{{\footnotesize $D$}}
\psfrag{ylabel}{{\footnotesize $C(D)$}}
\epsfxsize=3.in
\epsfclipon
\centerline{\epsffile{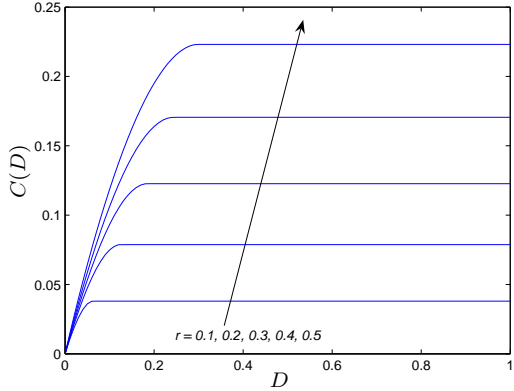}}
\caption{Capacity-distortion function for the scalar multiplicative channel.}
\label{fig:smultiply}
\end{figure}

After manipulations, we find that the optimal solution is given by
\begin{eqnarray*}
\mbox{If}&& D \geq r - \left[1 + e^{H_2(r)/r}\right]^{-1},\; p^\ast = \frac{1}{r}\left[1 + e^{H_2(r)/r}\right]^{-1},\\
&&\mbox{and}\;\; C(D) = H_2\left(p^\ast r\right) - p^\ast\cdot H_2(r);\\
\mbox{else}&& p^\ast = 1 - \frac{D}{r},\\
&&\mbox{and}\;\; C(D) = H_2(r - D) - \left(1 - \frac{D}{r}\right) H_2(r).
\end{eqnarray*}
From the solution, we observe the following. For relatively large $D$, the average distortion constraint is not active, and thus the optimal input distribution coincides with that for the unconstrained channel capacity. As the estimation distortion constraint $D$ falls below a threshold, the average distortion constraint becomes active, and the capacity-distortion function $C(D)$ deviates from the unconstrained channel capacity. We can show from the expression of $C(D)$ that, as $D \rightarrow 0$,
\begin{eqnarray}
\label{eqn:smultiply-smallD}
C(D) = \frac{\log(1 - r)}{-r} D + o(D).
\end{eqnarray}
Figure \ref{fig:smultiply} depicts $C(D)$ versus $D$ for different values of $r$. We notice that the tradeoff between communication rates and estimation distortions is evidently visible.

\subsection{A Block Multiplicative Channel}
\label{subsec:bmultiply}

A generalization of the scalar multiplicative channel is the following block multiplicative channel
\begin{eqnarray}
\underline{\rvy}_i = \rvs_i \underline{\rvx}_i,
\end{eqnarray}
where $\underline{\rvx}$ and $\underline{\rvy}$ are length-$K$ blocks so that the super-symbols in the block memoryless channel have alphabets $\mathcal{X}^K = \mathcal{Y}^K = \{0, 1\}^K$. The channel state $\rvs \in \mathcal{S} = \{0, 1\}$ remains fixed for each block, and changes in a memoryless fashion across blocks. We again adopt the Hamming distance as the distortion measure.

For such a channel, there are $2^K$ possible vectors for an input super-symbol. However, we note that, all of them except the all-zero $\underline{x} = \underline{0}$ are symmetric. This is because they all lead to the same conditional distribution for $\underline{\rvy}$ as well as the same minimum conditional distortion $d^\ast(\underline{x}) = 0$, $\forall \underline{x} \neq \underline{0}$. So from the concavity property of channel mutual information in input distributions, the optimal input distribution should take the following form:
\begin{eqnarray*}
P_{\underline{\rvx}}(\underline{0}) = 1 - p, \quad\mbox{and}\; P_{\underline{\rvx}}(\underline{x}) = p/(2^K - 1), \;\; \forall \underline{x} \neq \underline{0}.
\end{eqnarray*}
We can find that the channel mutual information per channel use is
\begin{eqnarray}
\frac{I(\underline{\rvx}; \underline{\rvy})}{K} = \frac{1}{K}\left\{H_2(pr) + p\cdot\left[r \log(2^K - 1) - H_2(r)\right] \right\},
\end{eqnarray}
and that the average distortion constraint is
\begin{eqnarray}
(1 - p)r \leq D,
\end{eqnarray}
the same as that in the scalar multiplicative channel case. After some manipulations, we find that the resulting optimal solution for general $K \geq 1$ is
\begin{eqnarray*}
\mbox{Case 1}&& 2^K > 1 + (1 - r)^{-1/r}:\\
&&p^\ast = 1, \; C(D) = \frac{r\log(2^K - 1)}{K}.\\
\mbox{Case 2}&& 2^K \leq 1 + (1 - r)^{-1/r}:\\
&&\mbox{if}\;\; D \geq r - \left[1 + \frac{1}{2^K - 1} e^{H_2(r)/r}\right]^{-1} \geq 0,\\
&& p^\ast = \frac{1}{r}\left[1 + \frac{1}{2^K - 1} e^{H_2(r)/r}\right]^{-1};\\
&&\mbox{else}\;\; p^\ast = 1 - \frac{D}{r}.\\
&& C(D) = \frac{1}{K} \left\{H_2(p^\ast r) + p^\ast \left[r \log(2^K - 1) - H_2(r)\right]\right\}.
\end{eqnarray*}

Case 1 arises because if the channel block length $K$ is sufficiently large such that $2^K > 1 + (1 - r)^{-1/r}$, then the resulting $p^\ast$ as given by Case 2 would be greater than one, which is impossible for a valid probability. In Case 1, we have $P_{\underline{\rvx}}(\underline{0}) = 0$, and all the nonzero symbols selected with equal probability $1/(2^K - 1)$.

In fact, Case 1 kicks in for rather small values of $K$. In our channel model we have assumed $r \in [0, 1/2]$. For $r$ smaller than $0.175$, Case 1 arises for $K \geq 2$; and for $r$ larger than $0.175$, Case 1 arises for $K \geq 3$.

In the scalar multiplicative channel ($K = 1$), we have noticed that $C(D)$ linearly scales to zero as $D \rightarrow 0$; see (\ref{eqn:smultiply-smallD}). For $K > 1$, however, we have
\begin{eqnarray}
C(0) = \frac{r\log(2^K - 1)}{K} > 0.
\end{eqnarray}
For comparison, let us consider a suboptimal approach based upon training that transmits $\rvx = 1$ in the first channel use in each channel block. The receiver can thus perfectly estimate the channel state $\rvs$ and achieve $D = 0$. The encoder then can use the remaining $(K - 1)$ channel uses in each channel block to encode information, and the resulting achievable rate is
\begin{eqnarray}
R(0) = \frac{r \log(2^{K - 1})}{K}.
\end{eqnarray}
Comparing $C(0)$ and $R(0)$, we notice that their ratio approaches one as $K \rightarrow \infty$, consistent with the intuition that training usually leads to negligible rate loss for channels with long coherence blocks.

\section{Further Issues}
\label{sec:issue}

In this section, we briefly discuss a few issues that are related to the capacity-distortion function formulation.

\subsection{Multiple Estimators and Other Cost Constraints}

In certain applications, multiple cost constraints may be present. For example, the receiver may be simultaneously interested in two or more different distortion measures, or the transmitter may have an average energy constraint for the channel input, besides the average distortion constraint. The multiple cost constraints should be simultaneously satisfied by augmenting the feasible set of input distributions, $\mathcal{P}_D$ (\ref{eqn:pd-defn}), to the intersection of multiple feasible sets, each for one cost constraint.

For either single or multiple cost constraints, the capacity-distortion function can be defined following Section \ref{sec:model}, formulated as a constrained channel coding problem following Section \ref{sec:formulation}, and computed following efficient algorithms like the Blahut-Arimoto algorithm \cite{blahut72:it, arimoto72:it} for discrete alphabets.

\subsection{Uncertainty in Channel State Statistics}

The constrained channel coding formulation in Section \ref{sec:formulation} can also be extended to the case in which the distribution of the channel state $\rvs$ is uncertain. For such a compound channel setting, we assume that the joint channel distribution $P_\theta(x, s, y) = P(y|x, s) P_\rvx(x) P_{\rvs, \theta}(s)$ is parametrized by an unknown parameter $\theta \in \Theta$, which is induced by the parametrized distribution of $\rvs$, $P_{\rvs, \theta}(s)$. If all the alphabets $\mathcal{X}, \mathcal{Y}$, and $\mathcal{S}$ are discrete, we can show following the proof in \cite{blackwell59:ams} that the capacity-distortion function of the compound channel is
\begin{eqnarray}
\sup_{P_\rvx \in \mathcal{P}_D} \inf_{\theta \in \Theta} I_\theta(\rvx; \rvy),
\end{eqnarray}
where
\begin{eqnarray}
\mathcal{P}_D = \left\{P_\rvx: \sum_{x \in \mathcal{X}} P_\rvx(x) d^\ast_\theta(x) \leq D, \forall \theta \in \Theta\right\}.
\end{eqnarray}
In $I_\theta(\rvx; \rvy)$ and $d^\ast_\theta(x)$, the subscript $\theta$ denotes that they are evaluated with respect to $P_\theta(x, s, y)$.

\subsection{Capacity Per Unit Distortion}

In light of the definition of channel capacity per unit cost for general cost-constrained channels \cite{verdu90:it}, we can analogously define the capacity per unit distortion, and show that it is equal to
\begin{eqnarray*}
C_d = \sup_{P_\rvx} \frac{I(\rvx; \rvy)}{\mathbf{E}[d^\ast(\rvx)]}.
\end{eqnarray*}
The capacity per unit distortion quantifies the maximum efficiency measured by the ratio between the amount of transmitted information and the incurred distortion in channel state estimation.

From \cite{verdu90:it}, if $d^\ast(x) = 0$ for at least two different input letters, then $C_d = \infty$; if there exists a unique $x_0 \in \mathcal{X}$ with $d^\ast(x_0) = 0$, then $C_d$ is also given by
\begin{eqnarray}
\label{eqn:cpud}
C_d = \sup_{x \in \mathcal{X}, x \neq x_0} \frac{D(P_{\rvy|x}\| P_{\rvy|x_0})}{d^\ast(x)},
\end{eqnarray}
where $D(\cdot\|\cdot)$ denotes the Kullback-Leibler divergence between two distributions. Here, note that in $P_{\rvy|\rvx}$ we marginalize over the channel state $\rvs$.

Given (\ref{eqn:cpud}), we can then conveniently evaluate $C_d$ for various channels. For example, the scalar multiplicative channel in Section \ref{subsec:smultiply} has $C_d = \frac{\log(1 - r)}{-r}$. In contrast, block multiplicative channels in Section \ref{subsec:bmultiply} with $K \geq 2$ have $C_d = \infty$, because all input letters except $\underline{0}$ lead to $d^\ast(\cdot) = 0$.

\section{Conclusions}
\label{sec:conclusion}

In this paper, we introduce a joint communication and channel estimation problem for state-dependent channels, and characterize its fundamental tradeoff by formulating it as a channel coding problem with input distribution constrained by an average ``estimation cost'' constraint. The resulting capacity-distortion function permits a systematic investigation of the channel property for communication and state estimation. Future research topics include specializing the general framework to particular channel models in realistic applications, and generalizing the results to multiuser systems and channels of generally correlated state processes.

\section*{Acknowledgment}
This work has been supported in part by NSF OCE0520324, the Annenberg Foundation, and the University of Southern California.

\bibliographystyle{ieee}
\bibliography{isit08}

\end{document}